\documentclass[journal]{IEEEtran}
\usepackage{cite}

\usepackage{color,soul}
\usepackage{gensymb}
\usepackage{dsfont}

\ifCLASSINFOpdf
  \usepackage[pdftex]{graphicx}
  \graphicspath{{../pdf/}{../jpeg/}}
  \DeclareGraphicsExtensions{.pdf,.jpeg,.png}
\else
  \usepackage[dvips]{graphicx}
  \graphicspath{{../eps/}}
  \DeclareGraphicsExtensions{.eps}
\fi
\usepackage[cmex10]{amsmath}
\usepackage{amssymb}

\usepackage{euscript}

\usepackage{multirow}
\usepackage{algorithm}
\usepackage[]{algpseudocode}
\usepackage{diagbox}
\usepackage{physics}
\usepackage{tikz}

\usepackage{amsthm}

\ifCLASSOPTIONcompsoc
  \usepackage[caption=false,font=normalsize,labelfont=sf,textfont=sf]{subfig}
\else
  \usepackage[caption=false,font=footnotesize]{subfig}
\fi
\usepackage{url}

\begin{document}
%
\title{A Hierarchical Deep Actor-Critic Learning Method for Joint Distribution System State Estimation}
%
%
%

\author{Yuxuan~Yuan,~\IEEEmembership{Graduate Student Member,~IEEE,}
	Kaveh~Dehghanpour,~\IEEEmembership{Member,~IEEE,}
	Zhaoyu~Wang,~\IEEEmembership{Member,~IEEE}
	and Fankun~Bu,~\IEEEmembership{Graduate Student Member,~IEEE,}
\thanks{This work is supported by the U.S. Department of Energy Office of Electricity Delivery and Energy Reliability under DE-OE0000875 (\textit{Corresponding author: Zhaoyu Wang}.)

 Y. Yuan, K. Dehghanpour, Z. Wang, and F. Bu are with the Department of
Electrical and Computer Engineering, Iowa State University, Ames,
IA 50011 USA (e-mail: yuanyx@iastate.edu; wzy@iastate.edu).
 }
}
%
%

\markboth{Submitted to IEEE for possible publication. Copyright may be transferred without notice}%
{Shell \MakeLowercase{\textit{et al.}}: Bare Demo of IEEEtran.cls for Journals}
%



\maketitle

\begin{abstract}
Due to increasing penetration of volatile distributed photovoltaic (PV) resources, real-time monitoring of customers at the grid-edge has become a critical task. However, this requires solving the distribution system state estimation (DSSE) jointly for both primary and secondary levels of distribution grids, which is computationally complex and lacks scalability to large systems. To achieve near real-time solutions for DSSE, we present a novel hierarchical reinforcement learning-aided framework: at the first layer, a weighted least squares (WLS) algorithm solves the DSSE over primary medium-voltage feeders; at the second layer, deep actor-critic (A-C) modules are trained for each secondary transformer using measurement residuals to estimate the states of low-voltage circuits and capture the impact of PVs at the grid-edge. While the A-C parameter learning process takes place offline, the trained A-C modules are deployed online for fast secondary grid state estimation; this is the key factor in scalability and computational efficiency of the framework. To maintain monitoring accuracy, the two levels exchange boundary information with each other at the secondary nodes, including transformer voltages (first layer to second layer) and active/reactive total power injection (second layer to first layer). This interactive information passing strategy results in a closed-loop structure that is able to track optimal solutions at both layers in few iterations. Moreover, our model can handle the topology changes using the Jacobian matrices of the first layer. We have performed numerical experiments using real utility data and feeder models to verify the performance of the proposed framework.

\end{abstract}


\section{Introduction}\label{sec:introduction}

As more stochastic customer-owned distributed resources, such as photovoltaic (PV) power generators, are connected to low-voltage (LV) secondary distribution grids, an urgent need grows for accurate system monitoring \cite{Zamani2014}. Specifically, topological details of secondary networks and the real-time measurements of customers have to be incorporated into distribution system state estimation (DSSE) to accurately capture voltage fluctuations across LV systems and quantify the impacts of these variations on medium-voltage (MV) primary distribution feeders. Recent years have seen a rapid growth in the deployment of smart meters, providing a good opportunity to achieve this \cite{Kaveh2019}.

\subsection{Literature Review and Challenges}
Most existing works have provided DSSE solutions only in a \textit{disjoint manner} (i.e., by decoupling primary and secondary networks); these works can be roughly categorized into two general groups: \textit{(1) Primary Grid DSSE:} various DSSE methods have been provided for MV primary distribution feeders, while aggregating all LV resources at the secondary transformers and disregarding the secondary grid topology and parameters \cite{Pegoraro2017,Pau2017,Wu2013,Therrien2013,Exposito2015,Hayes2015,Wakeel2016,YZ2019}. The basic approach is to compensate for lack of a detailed secondary model in DSSE by estimating LV network losses, which can then be used as pseudo-measurements to revise measurement aggregation \cite{Haughton2013}. \textit{(2) Secondary Grid DSSE:} Another group of papers have explored DSSE techniques for LV secondary networks while simplifying primary MV feeders \cite{Angioni2016,Huang2019,Majeed2012,Mutanen2013,Pertl2016,Bessa2018,Waeresch2015}. Here, the primary feeder has been generally modeled as a constant voltage source to which the secondary network is connected to.
\begin{figure*}[t]
      \centering
      \includegraphics[width=2\columnwidth]{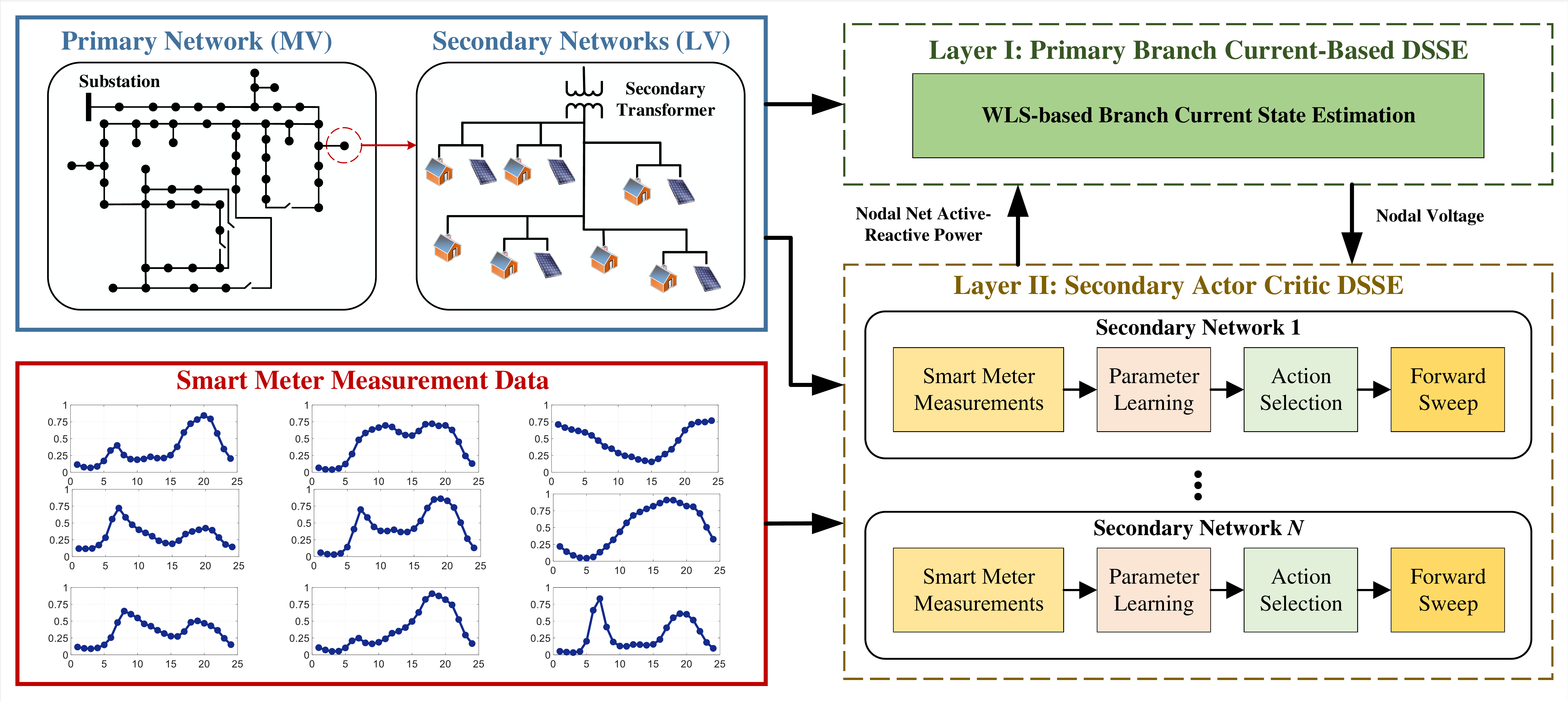}
\caption{Reinforcement learning-aided hierarchical DSSE framework.}
\label{fig:main}
\end{figure*}
Due to their disjoint approaches towards system monitoring, previous works in both groups can fail to accurately capture the potential mutual impacts of LV and MV networks on each other; furthermore, the mutual impacts of several neighboring secondary networks connected to the same primary feeder have not been quantified. Disjoint DSSE solvers become untenable and less accurate as conventional distribution systems move towards more active grids with higher penetration of renewable resources that can cause multi-directional power flow across the grid. Under this new situation, previous modeling assumptions, such as constant voltage level in primary feeders, can become too strong. Furthermore, the impact of secondary network topology on voltage fluctuations at the grid-edge can no longer be ignored. To address these shortcomings of previous works, it is imperative to devise a DSSE solution that is able to jointly monitor primary and secondary networks. Few recent papers \cite{Pau2018,Pau2019} have proposed distributed multi-level architectures for performing DSSE at LV and MV levels. However, in these cases the DSSE algorithms only have an open-loop one-directional flow of information from secondary to primary feeders which can fail to accurately capture the mutual impacts of LV-MV and LV-LV networks on each other, as the distribution grids become more active.

However, a joint approach towards DSSE faces several challenges: (1) a traditional weighted least squares (WLS) DSSE solver \cite{Kaveh2019} over a unified model of all primary and secondary circuits can lead to computational blow-up due to the extremely large size of joint primary-secondary systems, especially for urban systems. Traditional state estimation methods can take considerable time delays in practical systems, which does not truly reflect the current states \cite{YW2017}. This lack of scalability contributes to unacceptable time delays in obtaining system states and hinders online monitoring of distribution grids. (2) Accurate parameters of secondary networks, such as branch impedance values, can be unknown or highly uncertain in practice \cite{Peppanen2018}. This limits the ability of conventional WLS-based solvers to obtain reliable power flow-based Jacobian matrices. (3) Primary and secondary networks have distinct parametric characteristics. Compared to MV systems, the LV networks have higher R/X values and typical branch impedance levels. This characteristic difference between primary and secondary systems can lead to severe ill-conditioning of traditional joint DSSE solvers \cite{AM2012}. 

\subsection{Overall Structure of the Proposed Hierarchical Joint DSSE Framework}\label{sec:ml} 
To tackle these challenges, in this paper we have proposed a hierarchical reinforcement learning-aided framework for joint DSSE over primary and secondary distribution systems using  smart meter data, as shown in Fig. \ref{fig:main}. This framework consists of two layers: at the first layer, a WLS-based branch current state estimation (BCSE) algorithm is performed over the primary feeder to obtain the states of the MV distribution network, i.e., real/imaginary branch currents. At this layer, all the secondary circuits are treated as aggregated nodes with net equivalent active/reactive power injections provided by the second layer of the hierarchy. Since the WLS is performed only over the primary feeder, it is computationally efficient. After obtaining the states of the primary feeder, the solver passes down the estimated secondary transformer nodal voltages to the second layer of the hierarchy. 

At the second layer, a deep actor-critic (A-C) module \cite{Grondman2012} is trained for each LV network of secondary transformers. The basic idea behind the A-C approach is to estimate the states of secondary networks (i.e., secondary branch currents) to minimize the residuals of customer smart meter voltage measurements. Unlike WLS, the A-C modules leverage their past experiences to adaptively improve their future performance and generalize to unseen situations. Thus, the training process takes place offline and the A-C modules are employed online to estimate network states. Thanks to the neural network implementation of the A-C model, the online computation cost is several orders of magnitude lower than that of the WLS method. Furthermore, to estimate the secondary network states the A-C modules receive several inputs, including the latest WLS-based estimated secondary transformers' terminal voltages generated at the first layer and smart meter measurement data. Unlike the previous supervised learning-based works that only use the customer power data as the input of the learning model \cite{KR2019}, the voltage estimation at the service transformer is included as the input to capture the mutual impacts of LV and MV networks. The outputs of the second layer of the hierarchy, which are passed back to the first layer, are the net injected active/reactive powers to the primary feeder for each secondary transformer. These outputs are easily determined using the A-C-based estimated states of secondary circuits. Hence, the interaction between the two layers of the joint DSSE takes place at the secondary nodes, where nodal voltage flows from the first layer to the second layer and active/reactive power injections are passed in reverse. At each iteration of this closed-loop interaction, each layer revises the states of the network in response to the received inputs from the other layer.

Compared to previous works, our joint DSSE framework accurately quantifies the mutual impacts of primary-secondary networks, and secondary-secondary networks on each other, which is of practical importance in grids with high solar penetration. Furthermore, using the proposed A-C method, utilities can achieve a considerable speed-up in solving the joint DSSE in large-scale grids which allows them to monitor the whole system in near real-time. The distributed nature of the proposed framework allows for allocating the computational burdens of DSSE among multiple A-C modules, which further reduces the computation time. Moreover, compared to the traditional WLS-based method, our learning-based framework eliminates the need for pseudo-measurements to avoid the additional imputation error. Note that several previous papers have used supervised learning models to fully replace optimization solutions for estimating system states \cite{KR2019,WZ2020}. However, instead of approximating the joint distribution system state estimation directly with machine learning techniques, our method only approximates the secondary-level estimation process. Given that the secondary transformers are generally equipped with protection devices, the topology of the secondary distribution systems are constant\footnote{When an outage happens in a radial system, a protective device isolates the faulted area along with the loads downstream of the fault location (i.e., the whole secondary distribution systems}. Hence, the proposed method is able to handle the topology changes at the MV primary distribution networks using the Jacobin matrices in the first layer. Another advantage of the proposed hierarchical structure is that the A-C technique allows for explicit learning of the uncertainty of LV networks' states through parametric probabilistic policy functions. This not only helps operators to compensate for lack of accurate knowledge of secondary networks, but will also be integrated into the WLS formulation of the primary feeder to enhance overall monitoring accuracy. 

\begin{figure*}[ht]
      \centering
      \includegraphics[width=1.9\columnwidth]{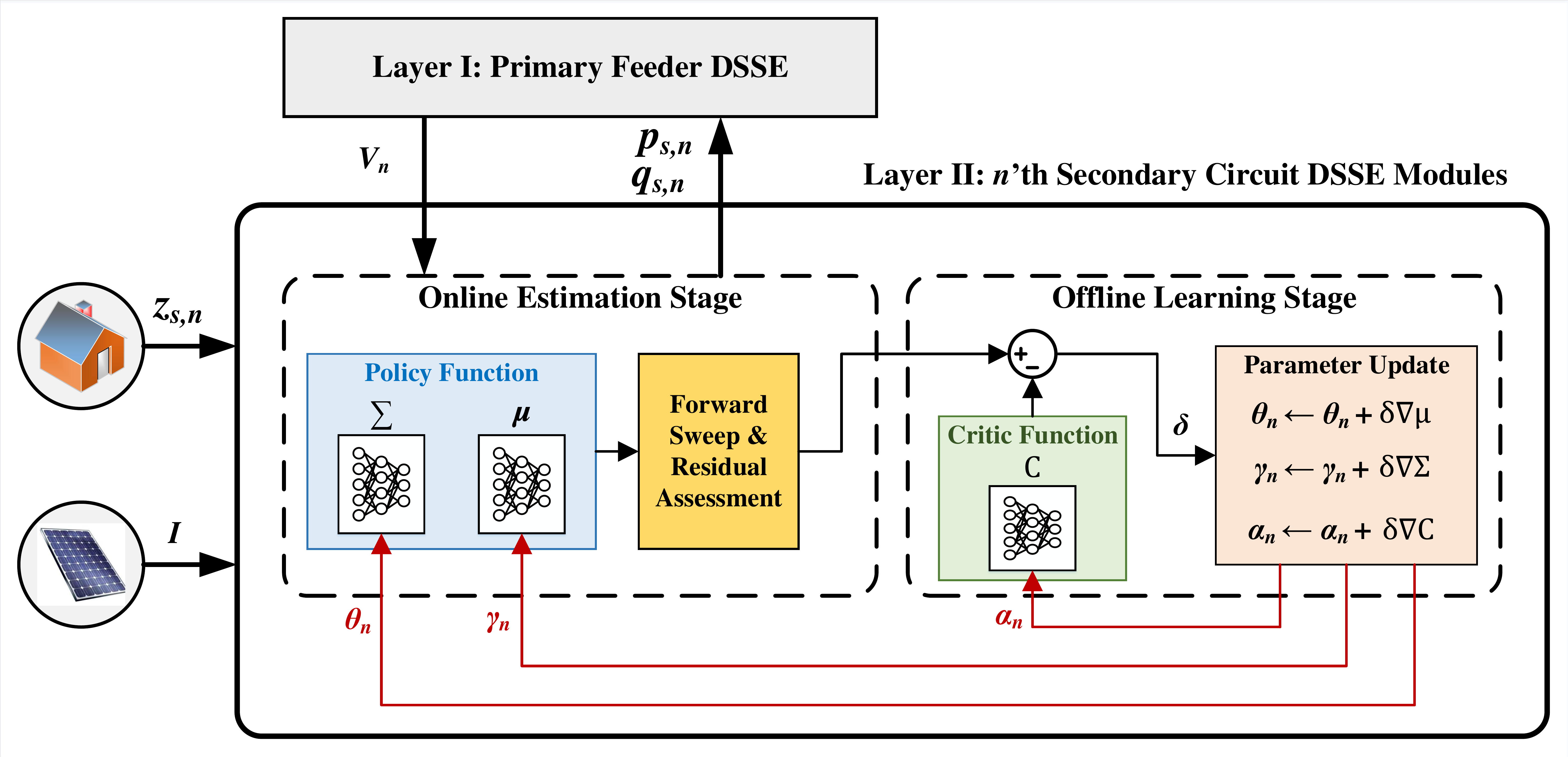}
\caption{Layer II: A-C-based DSSE for secondary circuits}
\label{fig:ac}
\end{figure*}

The rest of the paper is organized as follows: in Section \ref{sec:dsse}, the technical details of the proposed hierarchical joint DSSE are presented. In Section \ref{sec:numerical}, the numerical results have been analyzed to verify the performance of the joint DSSE method. In Section \ref{conclusion}, paper conclusions are presented.

\section{Deep Actor-Critic Strategy for Joint DSSE}\label{sec:dsse}

The goal of the proposed method is to provide an efficient model for improving distribution system situational awareness. In general, our joint DSSE model consists of two parts: an optimization-based solution that infers the system states of the primary-level network, and a learning-based method that estimates the customer-level states and provides a feedback to the first model. In the following, we describe the proposed model in detail.

\subsection{Primary Network BCSE}\label{sec:bcse}
At the first layer of the hierarchical joint DSSE, a WLS-based BCSE algorithm is performed over the MV network to minimize the sum of squared residuals ($J$) \cite{Baran1995}:
\begin{equation}
\label{eq:bcse0}
\begin{split}
\min_{\pmb{x_p}} J &= (\pmb{z_p}-\pmb{h}(\pmb{x_p}))^\top W (\pmb{z_p}-\pmb{h}(\pmb{x_p}))\\
&s.t. \ \ \ \pmb{z_p} = \left[
\begin{array}{c}
\pmb{z_{MV}} \\
\pmb{\hat{p}_s}\\
\pmb{\hat{q}_s}
\end{array}
\right]\\
&W = \left[
\begin{array}{ccc}
W_{MV} & \pmb{0} & \pmb{0}\\
\pmb{0} & W_{p_s} &\pmb{0}\\
\pmb{0}& \pmb{0} &W_{q_s}
\end{array}
\right]
\end{split}
\end{equation}
where, $\pmb{x_p}$ is a vector denoting the primary network states, including real/imaginary branch current values, $\pmb{z_p}$ is a vector containing the MV network sensor measurements ($\pmb{z_{MV}}$), including SCADA and $\mu$PMUs, and the estimated total active/reactive power injections of secondary transformers ($\pmb{\hat{p}_s}$, $\pmb{\hat{q}_s}$) that are provided by the second layer of the hierarchy. $\pmb{h}$ is the primary network measurement function which maps state values to measurements. Furthermore, $W$ is a weight matrix that represents the solver's confidence level in each element of $\pmb{z_p}$. $W$ consists of sub-matrces $W_{MV}$, $W_{p_s}$, and $W_{q_s}$ corresponding to $\pmb{z_{MV}}$, $\pmb{\hat{p}_s}$, and $\pmb{\hat{q}_s}$, respectively. Here, $W_{MV}$ is determined by the nominal accuracy levels of MV network sensors, e.g., the weight assigned to the measurements received from a specific sensor is selected as the inverse of measurement error variance for that sensor \cite{Baran1995}. The elements of $W_{p_s}$, and $W_{q_s}$ are determined by the estimated uncertainty of the secondary network states as elaborated in Section \ref{sec:a-c}.

Given the formulation \eqref{eq:bcse0}, the WLS-based solver performs the following steps to estimate the states of the primary network:
\begin{itemize}
\item\noindent\textbf{\textit{Step I:}} Receive the latest values of $\pmb{\hat{p}_s}$, $\pmb{\hat{q}_s}$, $W_{p_s}$, and $W_{q_s}$ from the second layer of the hierarchy (see Section \ref{sec:a-c}).

\item\noindent\textbf{\textit{Step II:}} Random state initialization ($\pmb{x_p}[0]$, $k\leftarrow 1$).

\item\noindent\textbf{\textit{Step III:}} At iteration $k$, update the measurement function \textit{Jacobian matrix}, $H$:
\begin{equation}
\label{eq:jac}
H = \frac{\partial\pmb{h}(\pmb{x_p}[k-1])}{\partial\pmb{x_p}}
\end{equation}
The elements of the Jacobian matrix for the BCSE method can be easily obtained for arbitrary feeders with known topology according to previous works \cite{Baran1995}.

\item\noindent\textbf{\textit{Step IV:}} Update the \textit{gain matrix}, $G$:
\begin{equation}
\label{eq:bcse3}
G(x) = H^\top(\pmb{x_p}[k-1])WH(\pmb{x_p}[k-1])
\end{equation}

\item\noindent\textbf{\textit{Step V:}} Update the state values using the gain and Jacobian matrices to reduce measurement residuals:
\begin{equation}
\label{eq:bcse4}
\pmb{x_p}[k] = \pmb{x_p}[k-1]+G^{-1}H^\top W (\pmb{z_p}-\pmb{h}(\pmb{x_p}[k-1]))
\end{equation}

\item\noindent\textbf{\textit{Step VI:}} $k\leftarrow k + 1$; go back to Step III until convergence, i.e., $||\pmb{x_p}[k] - \pmb{x_p}[k-1]||\leq \delta$, with $\delta$ being a user-defined threshold.

\item\noindent\textbf{\textit{Step VII:}} Given the estimated values of the branches, perform a forward sweep \cite{Baran1995} to obtain the voltages of secondary transformers throughout the network. Pass down the estimated voltage of the $n$'th secondary transformer ($V_n$) to the corresponding A-C module in the second layer of the joint DSSE hierarchy.
\end{itemize}

Compared to traditional state estimation solutions that use node voltages, BCSE adopts branch current as state variables, which is a more natural way of DSSE formulation for distribution systems \cite{Kaveh2019}. The simplification of the measurement functions helps improve the computation speed and memory usage. Therefore, BCSE is more suitable for large-scale distribution grids.

\subsection{Reinforcement Learning-Aided State Estimation for Secondary Networks}\label{sec:a-c}

The computational complexity of conventional WLS technique is determined by the matrix inversion, which induces a complexity of $O(N^3)$. Thus, running a BCSE algorithm over the whole primary and secondary networks at the same time is a computationally intensive task, especially for large-scale urban systems. To solve this challenge, the second layer of the hierarchy is designed with the objective of simplifying and speeding-up the joint DSSE process to achieve near real-time monitoring. Inspired by the recent success of machine learning techniques, we have leveraged a reinforcement learning technique, A-C method. Accordingly, an A-C module is trained for each secondary transformer as shown in Fig. \ref{fig:ac}. This module consists of two deep learning components that are trained cooperatively: (1) the \textit{actor} represents the secondary state estimation \textit{policy function} ($\pi_n$), which receives external inputs for the $n$'th secondary circuit, including the smart meter voltage/energy measurements ($\pmb{z}_{s,n}$), and the estimated transformer voltage from the first layer ($V_n$), and maps them to secondary states, $\pmb{x}_{s,n}$. Here, $\pmb{x}_{s,n}$ are the real/imaginary components of secondary circuit branch currents. To account for the uncertainty of the renewable resources in the state estimation process, this mapping is represented as a $D_n$-dimensional parametric multi-variate Gaussian probability distribution function, where $\pmb{x}_{s,n}\in \mathbb{R}^{D_n}$:
\begin{equation}
\begin{split}
&\pmb{x}_{s,n} \sim \pi_{n}(\pmb{\mu_n},\Sigma_n)\\
&=\frac{1}{\sqrt{|\Sigma_n|(2\pi)^{D_n}}}e^{-\frac{1}{2}(\pmb{x}_{s,n}-\pmb{\mu_n})^\top\Sigma_n^{-1}(\pmb{x}_{s,n}-\pmb{\mu_n})}
\end{split}
\label{eq:policy}
\end{equation}
where, $\pmb{c}_n = [\pmb{z}_{s,n}\ V_n]$, and $\pmb{\mu_n}$ and $\Sigma_n$ are the $n$'th secondary circuit state mean vector and covariance matrix, respectively. In this paper, these two statistical factors are parameterized using two deep neural networks (DNNs), $\mathcal{A}_\mu$ and $\mathcal{A}_\Sigma$, with parameters $\pmb{\theta_n}$ and $\pmb{\gamma_n}$: 
\begin{equation}
\pmb{\mu}_n=\mathcal{A}_\mu(\pmb{c}_n|\pmb{\theta_n})\label{DNN_1}
\end{equation}
\begin{equation}
\Sigma_n=\mathcal{A}_\Sigma(\pmb{c}_n|\pmb{\gamma_n})\label{DNN_2}
\end{equation}
Basically, parameters $\pmb{\theta_n}$ and $\pmb{\gamma_n}$ are the weight and biases assigned to the synapses in the DNNs, which need to be learned. This enables the operator to accurately quantify, not only the expected value of the secondary circuit states, but also their uncertainty, which is a critical element in grids with high renewable penetration. (2) The \textit{critic} is a DNN denoted by $\mathcal{C}$ with parameters $\pmb{\alpha_n}$ for the $n$'th circuit, which quantifies how well the actor is performing. In our problem, the critic tries to predict the secondary network DSSE residuals based on the inputs to the second layer:
\begin{equation}
\hat{r}_n=\mathcal{C}(\pmb{c}_{n}|\pmb{\alpha_n})\label{eq:critic_est}
\end{equation}
where, $\hat{r}_n$ represents the approximate residuals; ideally, if the critic has perfect performance, then, $\hat{r}_n = r_n$, meaning that the predicted residuals are equal to the realized measurement measurement residuals $r_n$.

Given the defined A-C modules, the computational process at the second layer of the hierarchy consists of an online state estimation stage (A), which is performed jointly with the first layer, and an offline parameter update stage (B), which is confined to the second layer alone:

\begin{itemize}
\item\noindent\textbf{\textit{Stage A - [Online Joint DSSE]}}

\item\noindent\textbf{\textit{Step A-I:}} Input the learned A-C parameters $\pmb{\theta_n}$, $\pmb{\gamma_n}$, and $\pmb{\alpha_n}$.

\item\noindent\textbf{\textit{Step A-II:}} Receive the updated $V_n$ from the first layer, and construct the external input vector, $\pmb{c}_n$.

\item\noindent\textbf{\textit{Step A-III:}} Construct the policy function $\pi_n$, according to \eqref{eq:policy}, using parameters $\pmb{\theta_n}$ and $\pmb{\gamma_n}$ and external inputs $\pmb{c}_n$.

\item\noindent\textbf{\textit{Step A-IV:}} Sample secondary circuit states in real-time using the constructed policy function, $\pmb{x}_{s,n} \leftarrow \pi_{n}$.

\item\noindent\textbf{\textit{Step A-V:}} Use generated states to perform a forward sweep \cite{Baran1995} over the secondary circuit to obtain the net active/reactive power injections at the transformer node, $\hat{p}_{s,n}$ and $\hat{q}_{s,n}$, as follows:
\begin{equation}
\hat{p}_{s,n} = V_nI_{Re,n}\label{ps}
\end{equation}
\begin{equation}
\hat{q}_{s,n} = V_nI_{Im,n}\label{qs}
\end{equation}
where, $I_{Re,n}\in \pmb{x}_{s,n}$ and $I_{Im,n}\in \pmb{x}_{s,n}$ are the estimated net real and imaginary components of $n$'th secondary transformer.

\item\noindent\textbf{\textit{Step A-VI:}} To construct $W_{p_s}$ and $W_{q_s}$, the variances of $\hat{p}_{s,n}$ and $\hat{q}_{s,n}$ need to be obtained. Noting that the uncertainty of LV circuits states are explicitly quantified by the covariance matrix of the policy function, $\pi_n$, we have:
\begin{equation}
\sigma^2_{p_{s,n}} = (V_n)^2\Sigma_{I_{Re,n}}\label{psvar}
\end{equation}
\begin{equation}
\sigma^2_{q_{s,n}} = (V_n)^2\Sigma_{I_{Im,n}}\label{qsvar}
\end{equation}
where, $\sigma^2_{p_{s,n}}$ and $\sigma^2_{q_{s,n}}$ are the variances of the net active and reactive power for the $n$'th LV system, and $\Sigma_{I_{Re,n}}$ and $\Sigma_{I_{Im,n}}$ are components of $\Sigma_n$ corresponding to the states $I_{Re,n}$ and $I_{Im,n}$, respectively. These variables are determined using $\mathcal{A}_\Sigma(\pmb{c}_n|\pmb{\gamma_n})$. Therefore, the weights assigned to $p_{s,n}$ and $q_{s,n}$ in the WLS-based solver of layer I are equal to $\sigma^{-2}_{p_{s,n}}$ and $\sigma^{-2}_{q_{s,n}}$, respectively.

\item\noindent\textbf{\textit{Step A-VII:}} Pass the net active/reactive power injection of all secondary transformers to the first layer of the joint DSSE framework, $\pmb{\hat{p}_{s}} = [\hat{p}_{s,1},...,\hat{p}_{s,N}]$ and $\pmb{\hat{q}_{s}} = [\hat{q}_{s,1},...,\hat{q}_{s,N}]$. Go back to Step A-II until $V_n$ is stabilized.
\begin{figure}
      \centering
      \includegraphics[width=1\columnwidth]{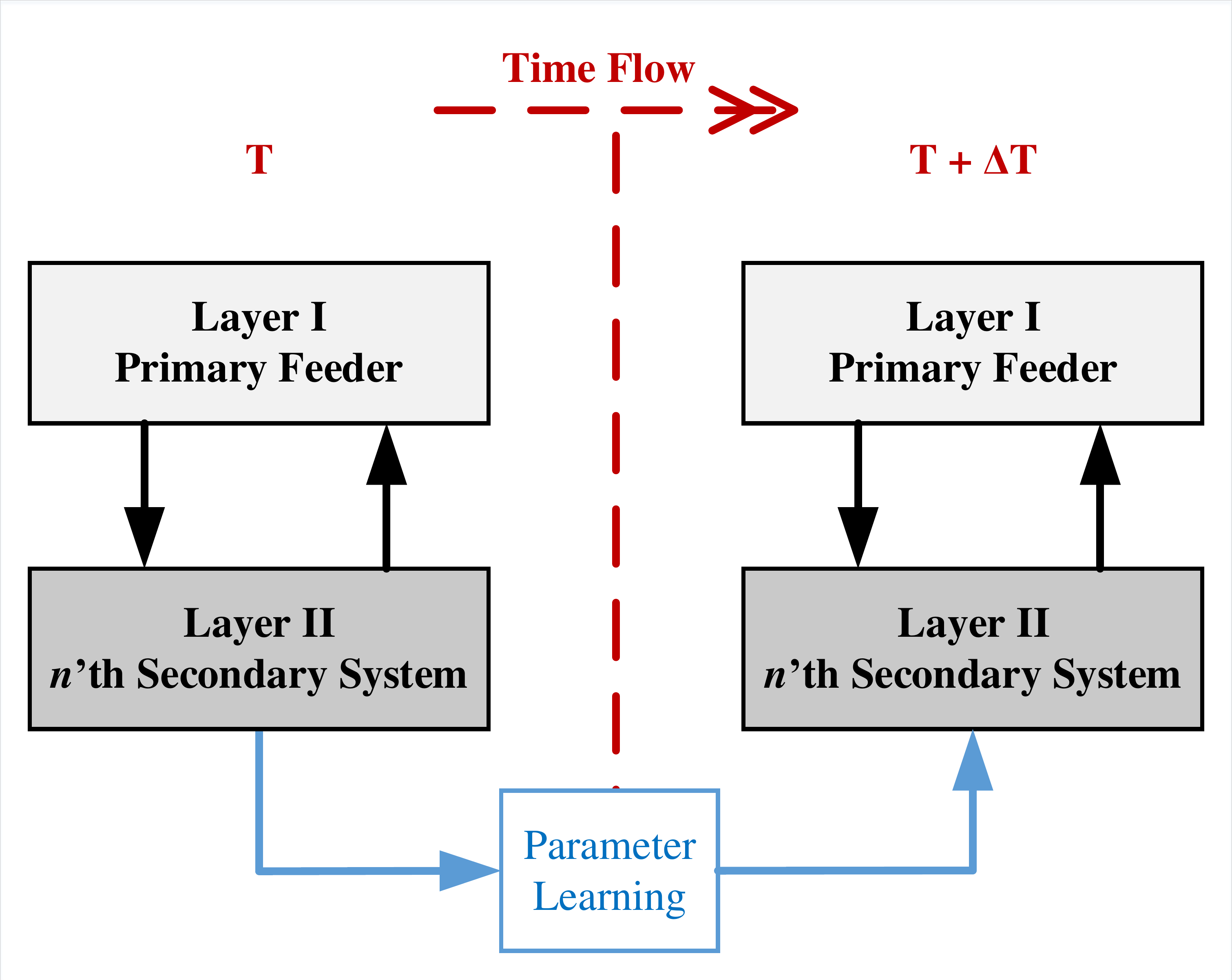}
\caption{Temporal function of the proposed hierarchical joint DSSE.}
\label{fig:temporal}
\end{figure}

\item\noindent\textbf{\textit{Stage B - [Offline A-C Parameter Update]}}

\item\noindent\textbf{\textit{Step B-I:}} After the online process has converged, re-sample states using the latest policy function, $\pmb{x}_{s,n} \leftarrow \pi_{n} + \pmb{u_n}$, where $\pmb{u_n}$ is a \textit{exploratory perturbation} generated using a zero-mean uniform distribution. This perturbation allows the A-C module to actively search for potential improvements in the learned policy and escape local minimums.

\item\noindent\textbf{\textit{Step B-II:}} Estimate the secondary DSSE residuals from the critic, using $\pmb{c}_{n}$ and DNN parameters $\pmb{\alpha_n}$, according to \eqref{eq:critic_est}.

\item\noindent\textbf{\textit{Step B-III:}} Use generated state sample and the latest value of $V_n$ from Step A-VII, to perform a forward sweep over the secondary circuit to obtain the estimated voltages; use the estimated nodal voltages to obtain the realized residual, $r_n$.

\item\noindent\textbf{\textit{Step B-IV:}} Obtain the \textit{temporal difference error} (TDE), $\delta_n = r_n - \hat{r}_n$, and use it to update the parameters of the critic:
\begin{equation}
\pmb{\alpha_n}\leftarrow \pmb{\alpha_n} + l_c \delta_n \nabla_{\pmb{\alpha_n}}\mathcal{C}(\pmb{c}_{n}) \label{eq:c-update}
\end{equation}
where, $l_c$ is a learning rate, and $\nabla_{\pmb{\alpha_n}}\mathcal{C}$ is the gradient of the critic DNN with respect to its parameters. This computation is performed using back-propagation over the DNN \cite{Grondman2012}.

\item\noindent\textbf{\textit{Step B-V:}} Update the parameters of the actor, using the TDE:
\begin{equation}
\pmb{\theta_n}\leftarrow \pmb{\theta_n} + l_{a} \delta_n\pmb{u_n}\nabla_{\pmb{\theta_n}}\pi_n(\pmb{c}_{n}) \label{eq:policyupdate}
\end{equation}
\begin{equation}
\pmb{\gamma_n}\leftarrow \pmb{\gamma_n} + l_{a} \delta_n\pmb{u_n}\nabla_{\pmb{\gamma_n}}\pi_n(\pmb{c}_{n}) \label{eq:policyupdate}
\end{equation}
with $l_{a}$ denoting the rate of policy learning. To obtain the gradient of policy function with respect to DNN parameters, $[\pmb{\theta_n},\pmb{\gamma_n}]$, chain rule is applied to the two sets of parameters separately:
\begin{equation}
\nabla_{\pmb{\theta_n}}\pi(\pmb{c}_{n}) = \frac{\Sigma_n^{-1}(\pmb{x}_{s,n}-\pmb{\mu}_n)}{\sqrt{|\Sigma_n|(2\pi)^{D_n}}}e^{-\frac{M}{2}}\nabla_{\pmb{\theta_n}}\mathcal{A}_\mu(\pmb{c}_{n}) \label{eq:actorupdate1}
\end{equation}
\begin{equation}
\begin{split}
&\nabla_{\pmb{\gamma_n}}\pi(\pmb{c}_{n}) =\\ &\frac{-\Sigma_n^{-1}(I-(\pmb{x}_{s,n}-\pmb{\mu}_n)(\pmb{x}_{s,n}-\pmb{\mu}_n)^\top\Sigma_n^{-1})e^{-\frac{M}{2}}}{2\sqrt{|\Sigma_n|(2\pi)^{D_n}}}\nabla_{\pmb{\gamma_n}}\mathcal{A}_\Sigma(\pmb{c}_{n}) 
\end{split}
\label{eq:actorupdate2}
\end{equation}
where, $M=(\pmb{c_n}-\pmb{\mu}_n)^\top\Sigma_n^{-1}(\pmb{c_n}-\pmb{\mu}_n)$ is an auxiliary matrix. Note that $\nabla_{\pmb{\theta_n}}\mathcal{A}_\mu$ and $\nabla_{\pmb{\gamma_n}}\mathcal{A}_\Sigma$ in \eqref{eq:actorupdate1} and \eqref{eq:actorupdate2} are obtained using back-propagation over the two DNNs of the actor.

\item\noindent\textbf{\textit{Step B-VI:}} Move to the next time-step; go back to Step A-I.
\end{itemize}

Fig. \ref{fig:temporal} shows the temporal functionality of the proposed A-C method. As can be seen, the parameters of DNNs are updated and replaced across time steps, while on the other hand, the bi-layer estimation takes place at each time step given the latest values of parameters. This enables the hierarchical framework to adapt to changes in the feeder across time, while offering fast real-time monitoring capability to utilities. 

\begin{figure*}[ht]
      \centering
      \includegraphics[width=2\columnwidth]{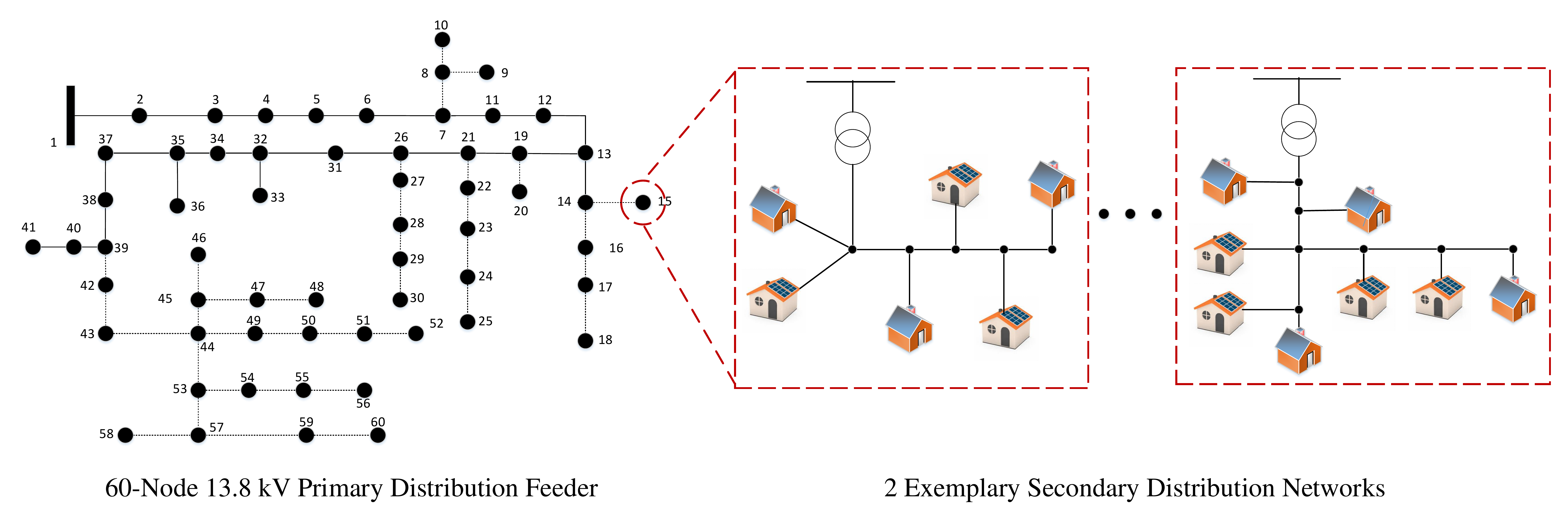}
\caption{Test feeder topology and secondary network examples}
\label{fig:testfeeder}
\end{figure*}

\subsection{Convergence Analysis}
The two layers of our model continuously exchange boundary information, including transformer voltages (first layer to second layer) and active/reactive total power injection (second layer to first layer). A major challenge in this model is to ensure the convergence of system monitoring, especially at the earlier stage of training when unreliable estimates generated by A-C modules may cause numerical instability for WLS. To avoid this, we have designed a confidence weight-based strategy. The basic idea is to integrate the TDE from the second layer (i.e., A-C modules) into the confidence matrix of the first layer (i.e., WLS). The TDE is able to measure how well the DNNs infer system states over time, which is a good metric for determining the reliability of the estimated secondary network states. Therefore, the A-C modules with lower TDE will receive higher confidence weights at the WLS. It should be noted that the A-C modules will be pre-trained using historical data, which further reduces the risk of numerical instability during online estimation.

\section{Numerical Results}\label{sec:numerical}
\begin{figure}
      \centering
      \includegraphics[width=1\columnwidth]{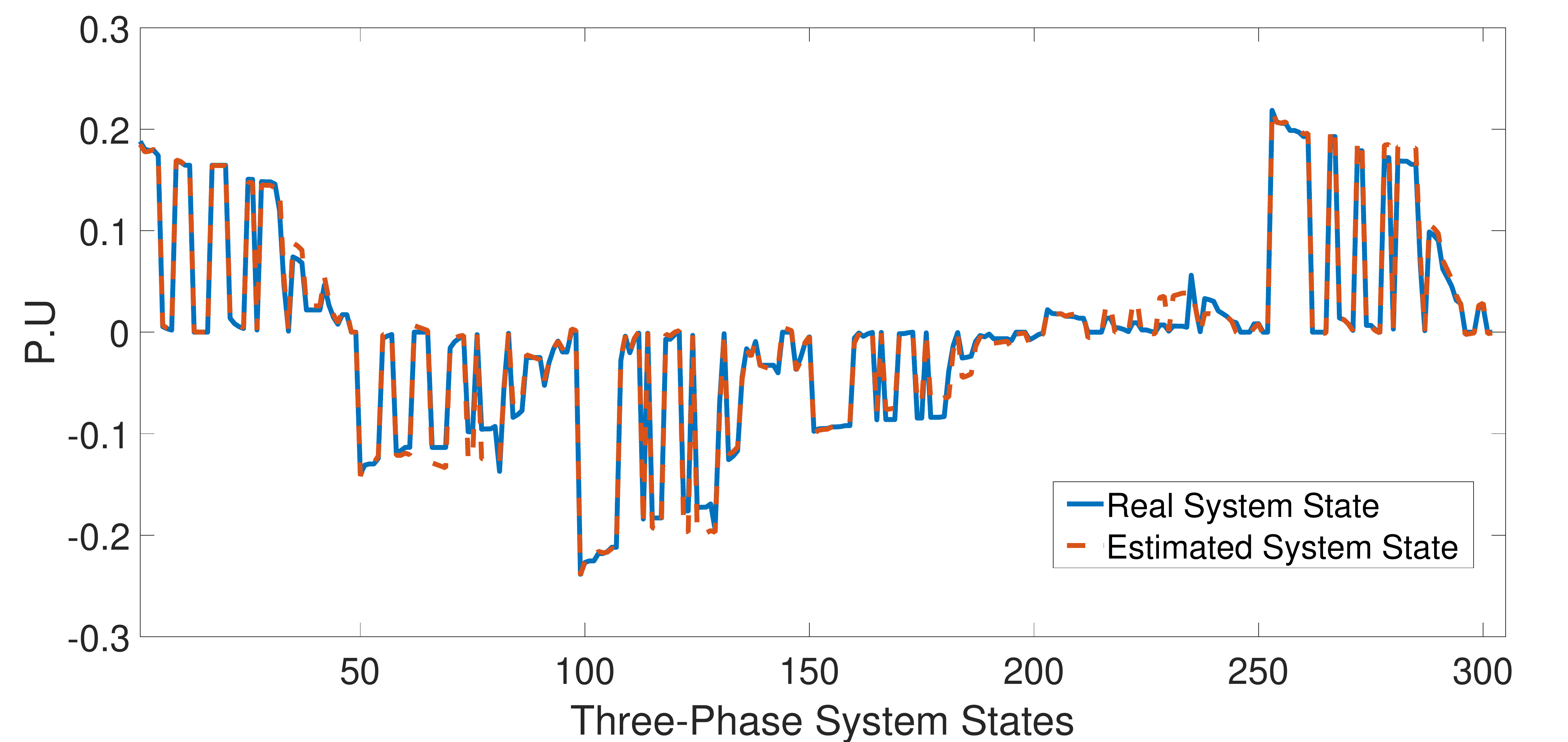}
\caption{Comparison of estimated three-phase system states and real values.}
\label{fig:primary_compare}
\end{figure}

\begin{figure}[htbp]
\centering
\subfloat  [Voltage magnitude component error]{
\includegraphics[width=3.35in]{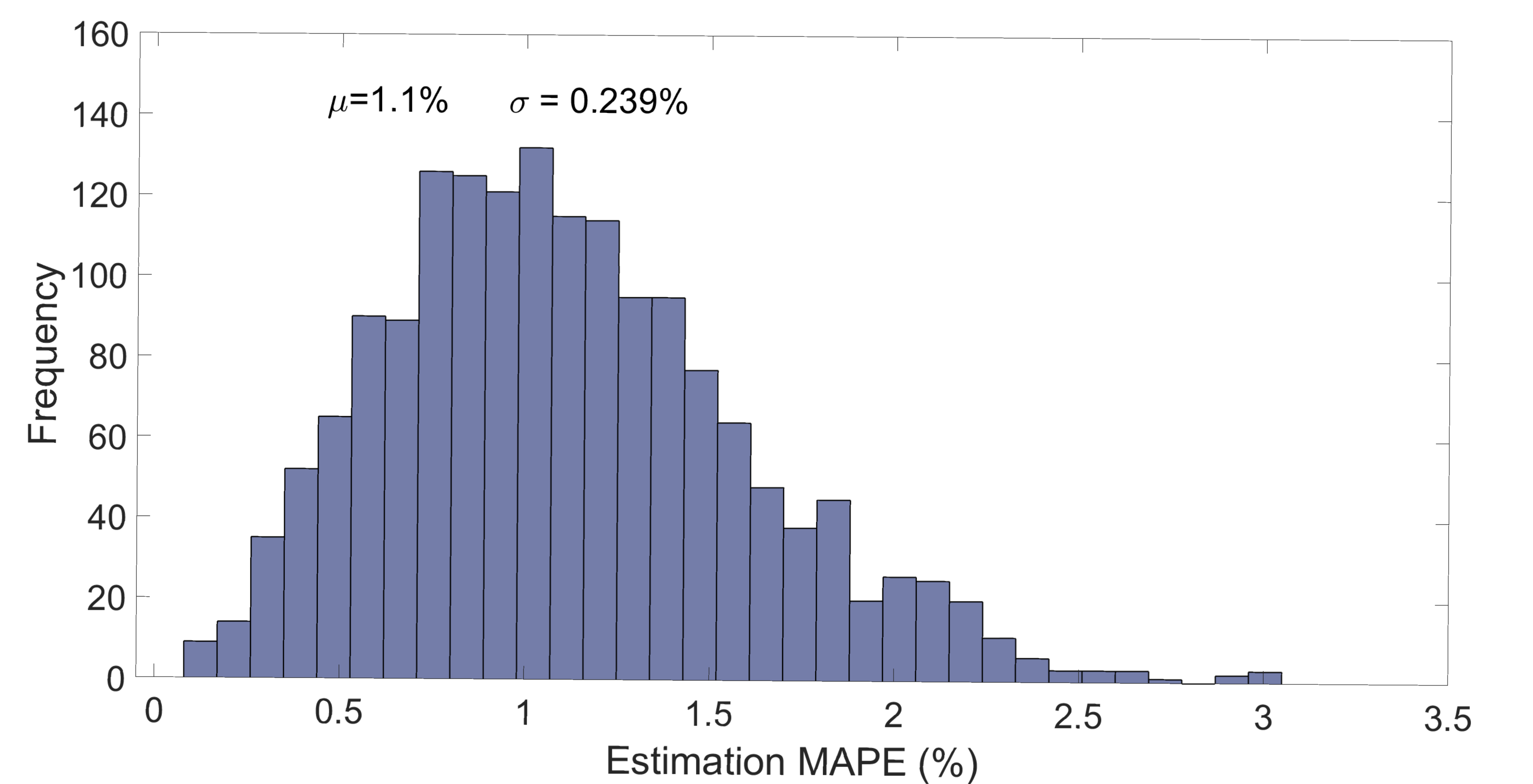}
}
\hfill
\subfloat [Voltage phase component error]{
\includegraphics[width=3.35in]{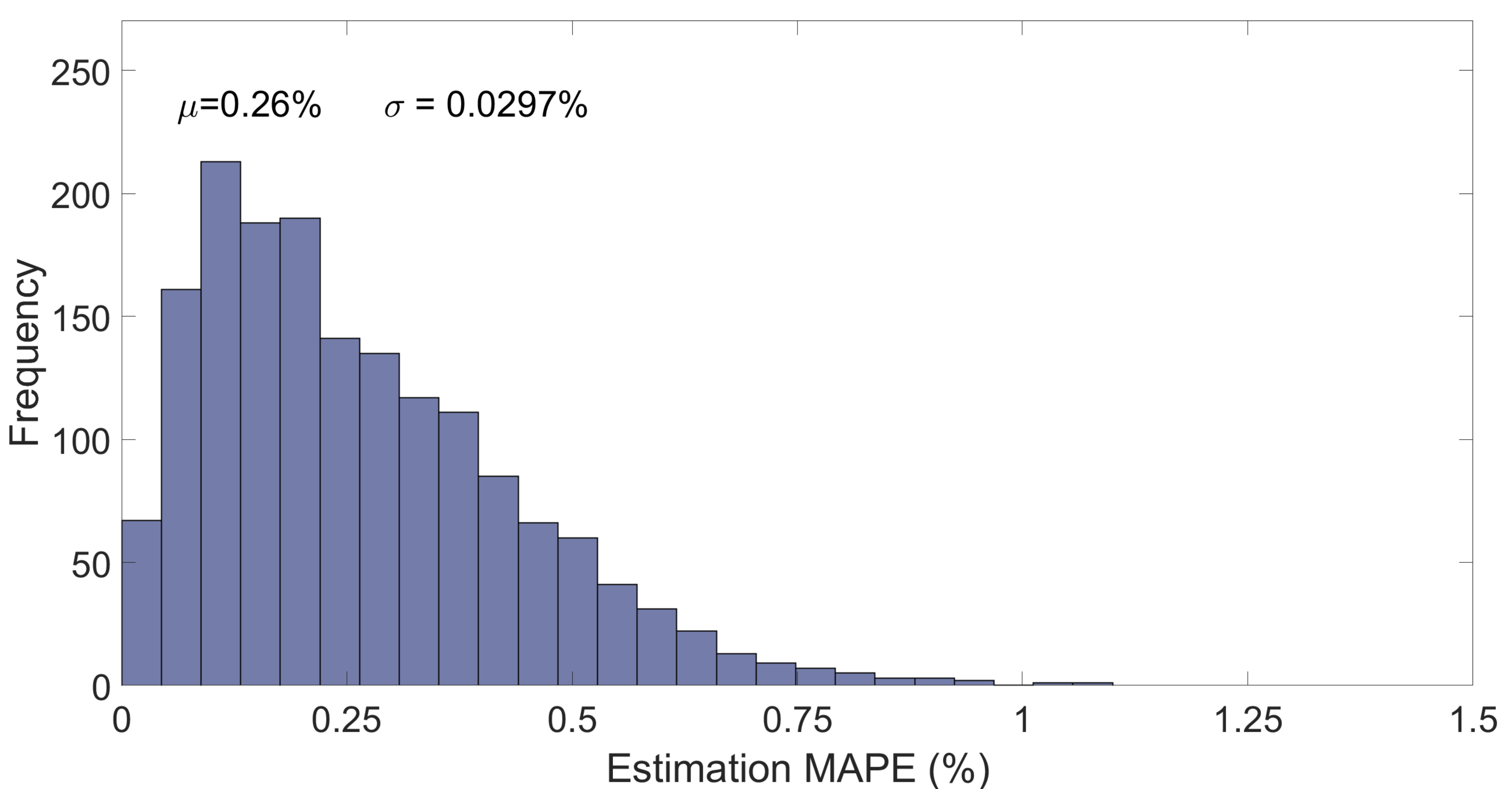}
}
\caption{State estimation performance using the proposed reinforcement learning-aided hierarchical DSSE model.}
\label{fig:error_hist}
\end{figure}
The proposed joint DSSE framework is verified using real SCADA/smart meter data and MV-LV network OpenDSS models from an utility partner in the U.S. The data includes customers' energy/voltage measurements at the secondary networks, and total primary feeder active/reactive power and substation voltages. More details on the data are available online \cite{Test_system}. Since we have real smart meter data for each customer, the transformer-level load consumption is aggregated based on the corresponding customer-level smart meter data. The test system consists of a 60-node 13.8 kV primary feeder and 44 secondary circuits with a total number of 238 customers, as shown in Fig. \ref{fig:testfeeder}. In this figure, we have shown the topology of the primary feeder and two exemplary secondary networks. Distributed solar resources are added to the secondary networks to capture the impact of uncertain renewable resources on DSSE. The penetration level of renewable power is $50\%$ with respect to the long-term average peak load. The solar power data is adopted from \cite{solar}. To validate our hierarchical reinforcement learning-aided DSSE framework, we have assumed that $10\%$ of customers in this feeder have smart meters. To build the training dataset for A-C method, the consumption data for the rest of unobservable customers are randomly generated. The corresponding system states are obtained using the OpenDSS. In DSSE, the maximum error values for the real measurements is $3\%$. In this work, the hyperparameter set of the A-C modules are calibrated by using the random search strategy \cite{randomsearch1}. As a result, the three DNNs, $\mathcal{A}_\mu$, $\mathcal{A}_\Sigma$, and $\mathcal{C}$, consist of 3 hidden layers of 10 neurons. The learning rates of actor and critic, $l_a$ and $l_c$, are selected as 0.01.

The A-C module is trained for various secondary networks in parallel based on the historical/simulation data\footnote{When the system is fully observable (i.e., smart meters are installed for each customer), we use the historical data to train our model. If the system is partially observable, a parametric model is used to learn the underlying marginal distribution of the customer load consumption and then generate a training set based on this distribution. The OpenDSS software is leveraged to obtain the system states.} and tested using the new data inquiry. After model training, Fig. \ref{fig:primary_compare} compares the estimated primary-level distribution system states (i.e., branch current real and imaginary parts) with the actual state values using the proposed method at a specific time point. As is demonstrated in the figure, the outcome of our method closely follows the underlying states. It should be noted that our test network is a three-phase unbalanced distribution system and the phase connections of customers are known. Furthermore, to validate the average performance of the proposed method, we have tested our method over a long-term period (more than 1500 time points). The error distribution of state estimation is shown in Fig. \ref{fig:error_hist}. The Mean Absolute Percentage Error (MAPE) criterion is used here to evaluate the accuracy of state estimation:
\begin{equation}
\label{eq:MAPE}
M = \frac{100\%}{n_s}\sum_{t=1}^{n_s}|\frac{\hat{A(t)}-A(t)}{\hat{A(t)}}|
\end{equation}
where, $\hat{A(t)}$ and $A(t)$ are the actual state value and the estimated value. As is demonstrated in these figures, the estimation errors for voltage magnitude and phase angle are $1.1\%$ and $0.26\%$, respectively. These results corroborate the satisfactory performance of the proposed model over real data.
\begin{figure}
      \centering
      \includegraphics[width=1\columnwidth]{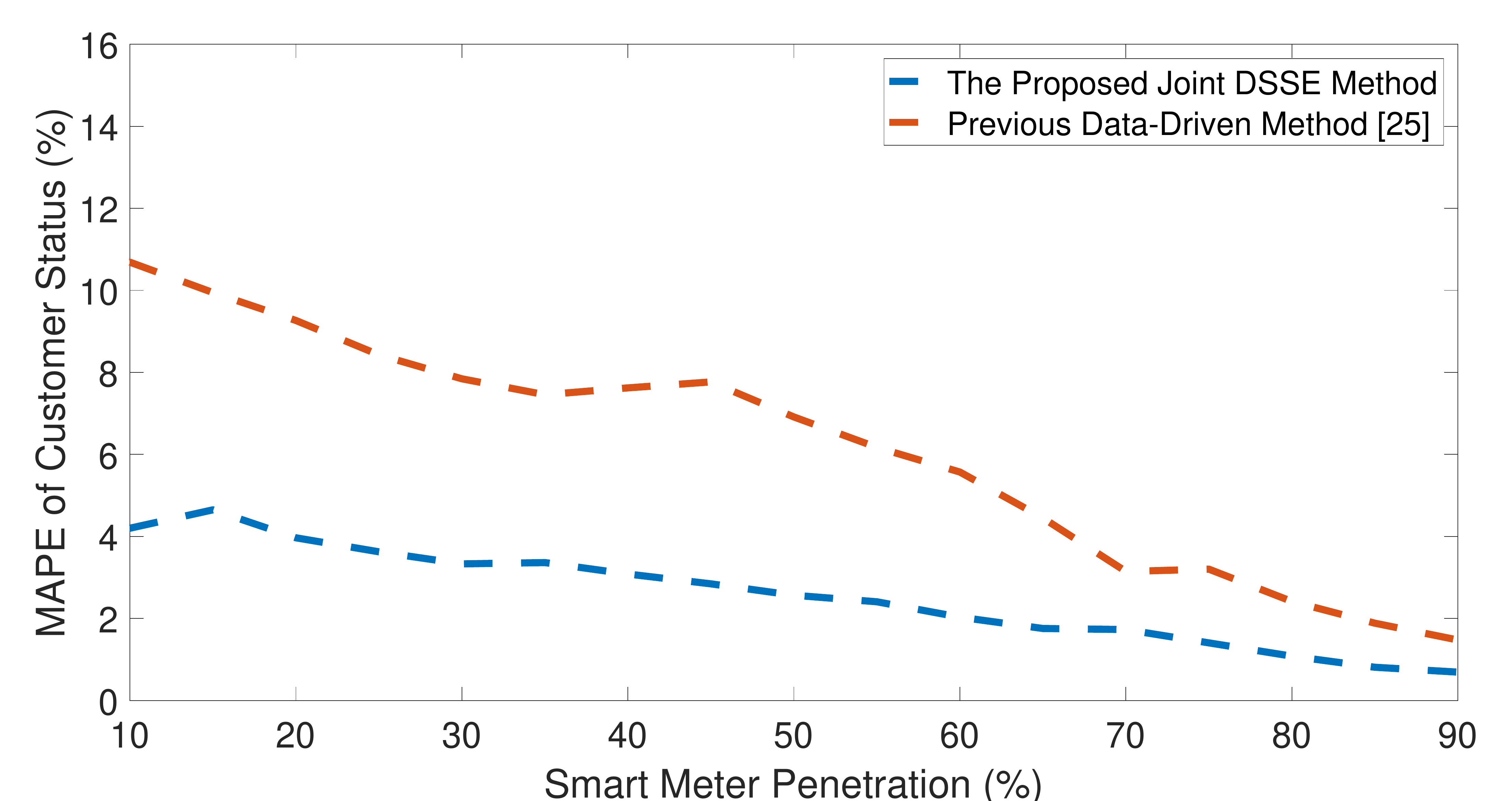}
\caption{Sensitivity analysis: quantifying the impact of observability on estimation accuracy.}
\label{fig:sensitivity}
\end{figure}
\begin{figure}[tbp]
\centering
\subfloat [Probability density function of online action selection time]{
\includegraphics[width=1\columnwidth]{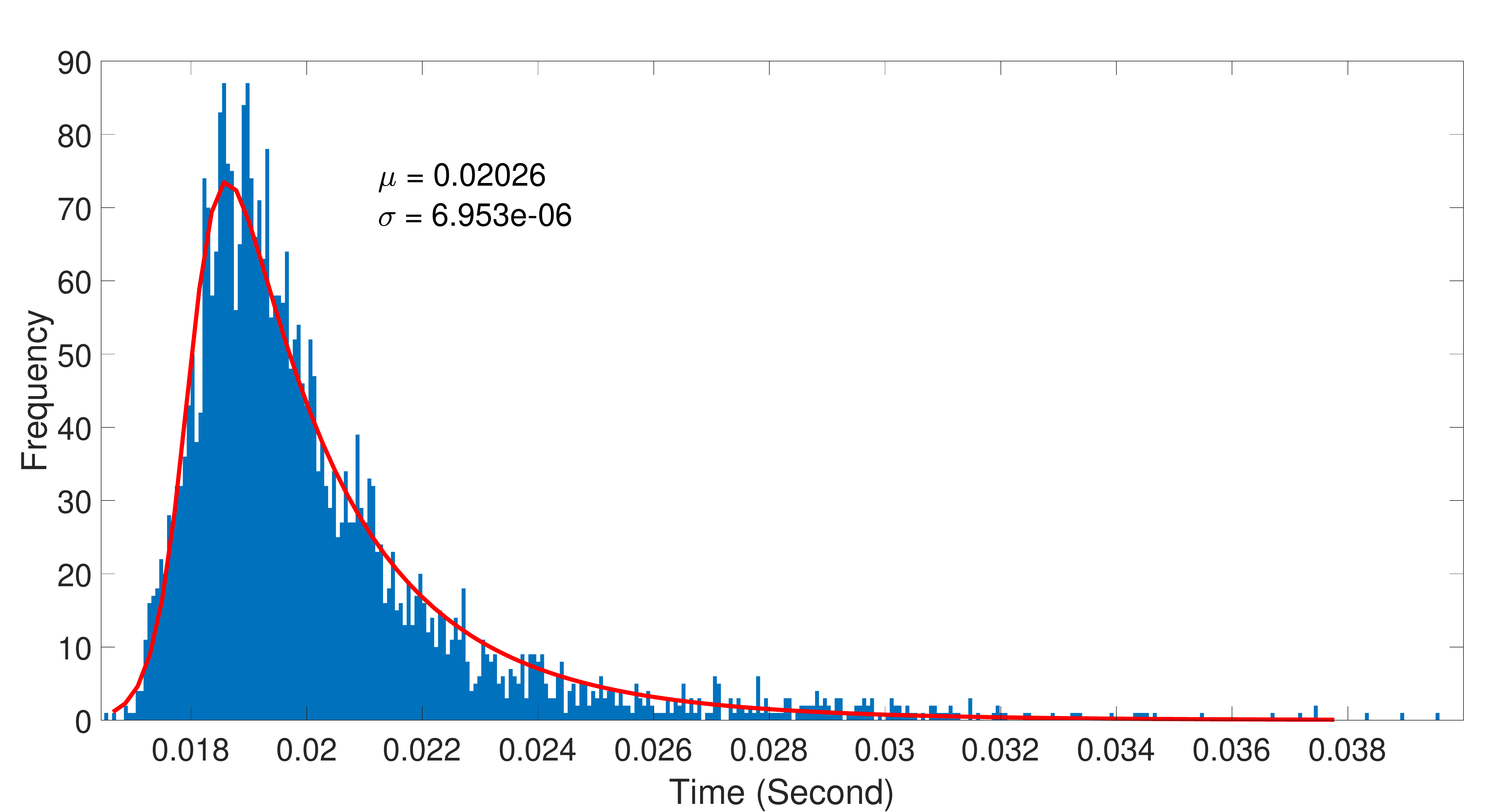}
}
\hfill
\subfloat [Cumulative distribution function of online action selection time]{
\includegraphics[width=1\columnwidth]{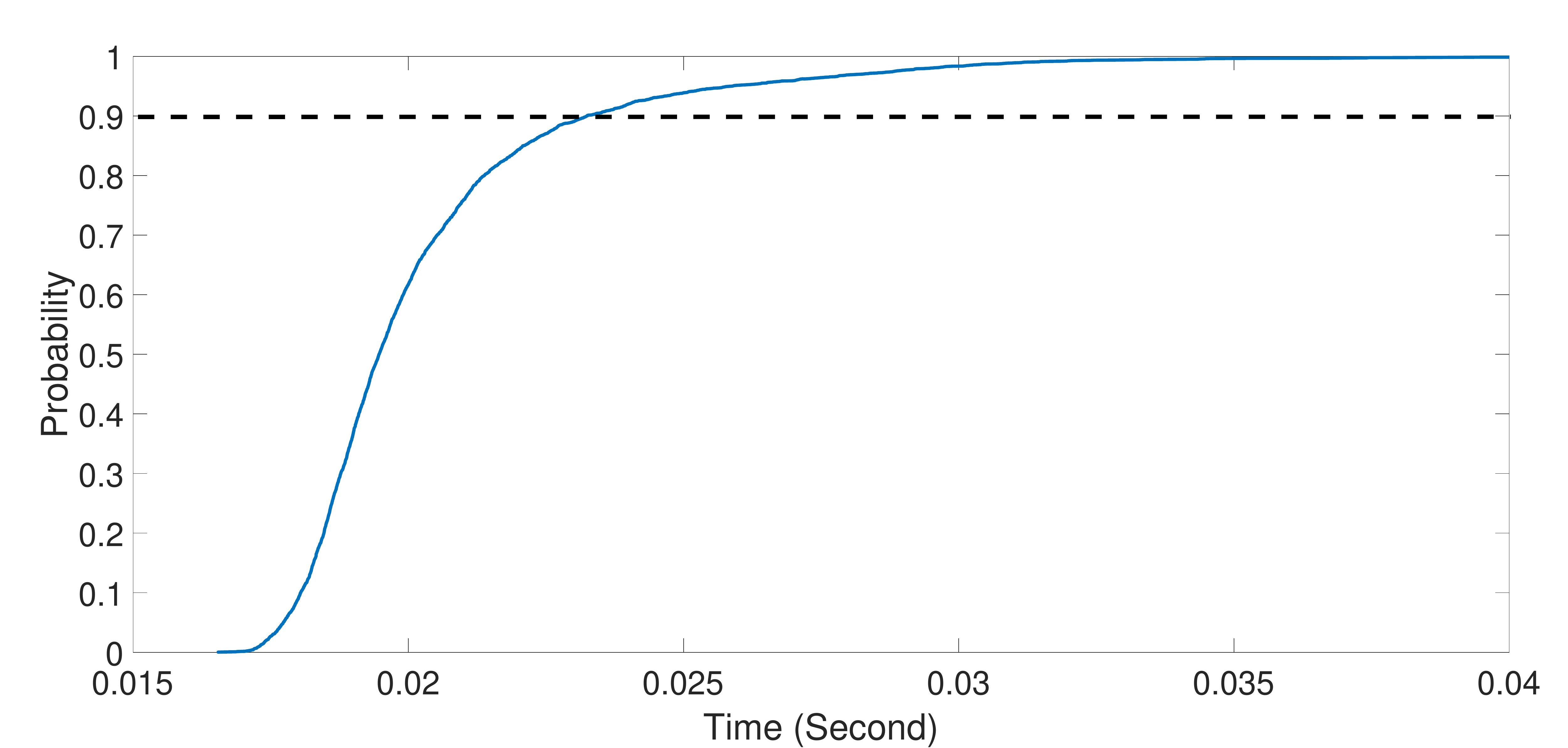}
}
\caption{Statistical results of online action selection time.}
\label{fig:oneline_time}
\end{figure}

Even though the learning-based method can eliminate the need for pseudo-measurement generation, the system observability (i.e., smart meter penetration) still impacts its performance. To represent the sensitivity of the state estimation accuracy to the system observability, Fig. \ref{fig:sensitivity} shows the secondary-level state estimation accuracy of the proposed method under various smart meter penetration levels. Moreover, in this figure, the accuracy of a previous learning-based method is compared with our solution \cite{KR2019}. It can be observed that the estimation accuracy decreases as the percentage of smart meter penetration decreases due to the information loss. Thanks to its hierarchical nature, our method outperforms the existing learning method at all observability levels.

\begin{figure}
      \centering
      \includegraphics[width=1\columnwidth]{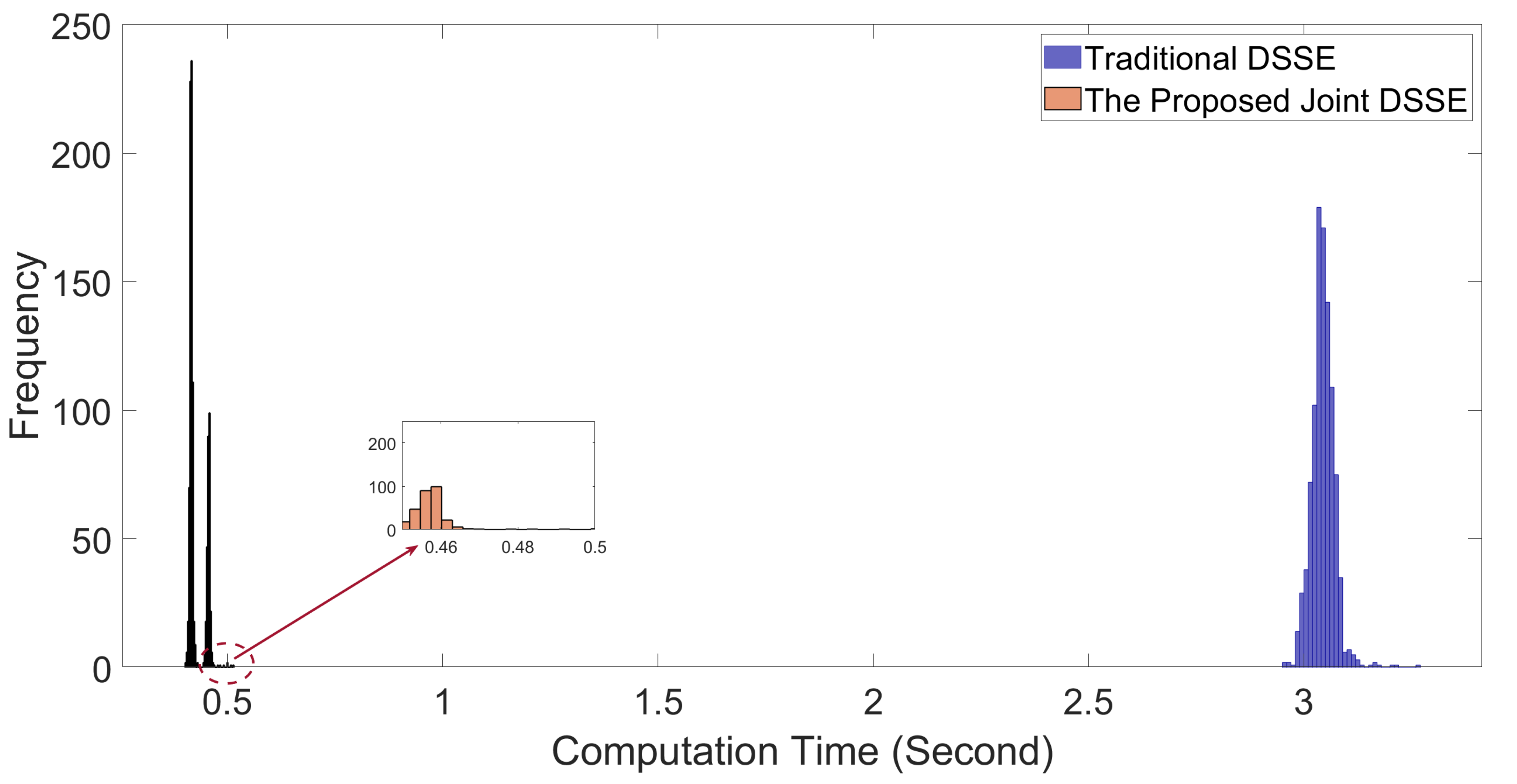}
\caption{Computation time comparisons (the proposed method versus a previous method).}
\label{fig:time}
\end{figure}

To ensure that the proposed method can provide real-time monitoring in practice, we have tracked the computation time. Note that the case study is conducted on a standard PC with an Intel(R) Xeon(R) CPU running at 3.70 GHz and with 32.0 GB of RAM. Fig. \ref{fig:oneline_time} presents the computation time distribution of the online action selection of A-C modules. Considering the uncertainty of the computation speed, $3500$ Monte Carlo simulations have been performed. As shown in the figure, the majority of online action time are concentrated around $0.02$ second. Moreover, based on the cumulative distribution function of online action time, almost $90\%$ of simulations have online action time below $0.024$ seconds, thus ensuring real-time system monitoring. Moreover, the computation time of the whole hierarchical framework is tested and compared to the traditional DSSE method \cite{Wang2004}. Fig. \ref{fig:time} shows the computation time distributions of our proposed method and an existing monitoring model \cite{Wang2004} over a 60-node distribution network. In this case, our framework is able to significantly improve the computation time by an average factor of $6$ times. It should be noted that our test system is a middle-size rural distribution feeder that has a limited number of customers. Since the computation burden of the optimization method grows exponentially, our method's improvements in computation time would be higher in large-scale urban systems.

\section{Conclusion}\label{conclusion}
In this paper, we have presented a reinforcement learning-aided hierarchical DSSE solution to jointly monitor the primary and secondary networks. Compared to previous works, the proposed solution is scalable to large grids and can accurately capture the impact of volatile grid-edge renewable resources on system states. Our model enables fast online estimation of secondary network states, while allowing for offline evaluation and updates of DNNs. The hierarchical joint DSSE method has been tested using real smart meter data and models of distribution grids. It is observed that after the estimation policy function is fully learned, the proposed method can accurately estimate the primary and secondary system states. Moreover, the results show that this solution is able to outperform previous monitoring methods in terms of estimation accuracy and computation time.

\ifCLASSOPTIONcaptionsoff
  \newpage
\fi



\bibliographystyle{IEEEtran}
\bibliography{IEEEabrv,./bibtex/bib/IEEEexample}
\end{document}